\documentclass[sigconf]{acmart}

\AtBeginDocument{%
\providecommand\BibTeX{{%
\normalfont B\kern-0.5em{\scshape i\kern-0.25em b}\kern-0.8em\TeX}}}

\usepackage{steinmetz}  
\usepackage{url}
\begin{document}

%
\title[Small-Signal Stability Analysis: Droop-Controlled Microgrids]{Small-Signal Stability Analysis for Droop-Controlled Inverter-based Microgrids with Losses and Filtering}

\author{Abdullah Al Maruf}
\affiliation{%
  \institution{Washington State University}}
\email{abdullahal.maruf@wsu.edu}

\author{Mohammad Ostadijafari}
\affiliation{%
  \institution{Washington State University}}
\email{m.ostadijafari@wsu.edu}

\author{Anamika Dubey}
\affiliation{%
  \institution{Washington State University}}
\email{anamika.dubey@wsu.edu}

\author{Sandip Roy}
\affiliation{%
  \institution{Washington State University}}
\email{sandip@wsu.edu}

%
\renewcommand{\shortauthors}{Maruf, Ostadijafari, Dubey and Roy}

%

\begin{abstract}
An islanded microgrid supplied by multiple distributed energy resources (DERs) often employs droop-control mechanisms for power sharing. Because microgrids do not include inertial elements, and low pass filtering of noisy measurements introduces lags in control, droop-like controllers may pose significant stability concerns. This paper aims to understand the effects of droop-control on the small-signal stability and transient response of the microgrid. Towards this goal, we present a compendium of results on  the small-signal stability of droop-controlled inverter-based microgrids with heterogeneous loads, which distinguishes: (1) lossless vs. lossy networks; (2) droop mechanisms with and without filters, and (3) mesh vs. radial network topologies. Small-signal and transient characteristics are also studied using multiple simulation studies on IEEE test systems. 
\end{abstract}
%
%



%

\copyrightyear{2019} 
\acmYear{2019} 
\setcopyright{acmcopyright}
\acmConference[e-Energy '19]{Proceedings of the Tenth ACM International Conference on Future Energy Systems}{June 25--28, 2019}{Phoenix, AZ, USA}
\acmBooktitle{Proceedings of the Tenth ACM International Conference on Future Energy Systems (e-Energy '19), June 25--28, 2019, Phoenix, AZ, USA}
\acmPrice{15.00}
\acmDOI{10.1145/3307772.3328310}
\acmISBN{978-1-4503-6671-7/19/06}


\maketitle

	\section{Introduction}
    \label{sec:Introduction}
     A microgrid is an interconnected low/medium-voltage power distribution network primarily supplied by inverter-based distributed energy resources (DERs) and can operate in both grid connected and islanded modes \cite{ashabani2012flexible,bouzid2015survey}. To enable a stable and economical operation, a hierarchical control framework comprised of primary, secondary and tertiary controllers operating at different time-scales is often employed \cite{guerrero2011hierarchical,guerrero2013advanced}. In islanded-mode, the primary controller is responsible for maintaining a stable grid operation and often utilizes a droop controller that mimics the inertial response of a synchronous generator for a highly inductive grid \cite{chandorkar1993control, visscher2008virtual}. As it is mentioned in \cite{pogaku2007modeling}, increasing the droop gains improves the power sharing while it adversely affects the overall system stability. Despite this inherent trade-off between power sharing and stability, the ability to distribute the total demand among DERs using local feedback signals, without the need for communications, makes the droop-based controllers an appropriate choice for providing a fast response \cite{farrokhabadi2017frequency}. However, because microgrids do not contain inertial elements and controls involve fast, noisy measurements, droop-like controllers implemented in DERs may pose significant stability concerns, and hence require a comprehensive stability analysis \cite{vandoorn2013microgrids}. 

Small-signal stability analyses of islanded microgrid dynamics using linearized models have been recently considered in the literature \cite{ainsworth2013structure,song2015small,simpson2013synchronization,song2017network,schiffer2013synchronization,monshizadeh2018novel}. The analyses that evaluate network-level properties often draw on  algebraic-graph-theory formalisms, which provide useful constructs for the dynamic analysis of interconnected systems. In particular, algebraic graph theory has been used to understand the relationship between a droop-controlled microgrid network's structure and its dynamical behavior \cite{simpson2013synchronization,ainsworth2013structure,song2015small}. For example, \cite{simpson2013synchronization}, presents a model for lossless inductive microgrid with frequency-droop control similar to Kuramoto model of phase-coupled oscillators, and derives conditions for obtaining a stable and synchronous solution for the network. Building on \cite{simpson2013synchronization}, \cite{ainsworth2013structure} proposes structure-preserving models for lossless frequency-droop controlled inverter-based microgrid and derives the necessary condition for frequency synchronization. The concept of the active power flow graph is introduced in \cite{song2015small} where the authors prove that small-signal stability is equivalent to the positive semi-definiteness of the resulting Laplacian matrix. 

The algebraic graph theory formalism developed in \cite{simpson2013synchronization,ainsworth2013structure,song2015small} provides a foundation for the modeling and stability analyses of droop-controlled microgrids. The models in the aforementioned literature, however, do not represent the lag/sluggishness in droop controllers that is either inherently present, or deliberately introduced due to low-pass filters employed to decrease the measurement noise in controller's input variables. In \cite{liu2016comparison}, the authors recognize that the delay introduces an inertial response that makes the frequency change at a slower rate when compared to the changes in active power flow, however, a formal network-level analysis is not undertaken. A model for internal inverter control together with output filter is developed in \cite{kotpalliwar2015modelling}. However, these studies do not determine the dependence of the dynamics on the network topology and the filter parameters introducing first-order lag. Also, they assume a lossless highly inductive microgrid, and do not investigate the effects of time-lagged droop-control on the stability of microgrid with lossy lines. 

Although several theoretical frameworks have been proposed to understand the angle stability of the droop-controller microgrids, there remain a critical gap with regard to theoretical stability guarantees in the presence of filtering in droop control and/or power system losses due to resistive lines. Both filtering and resistive/lossy lines are practical realities for a microgrid and cannot be neglected in any stability argument. To address this critical gap in the existing literature, in this paper, we present a compendium of results stating graph-theoretic conditions to ensure angle stability of inverter-based microgrids in the small-signal sense with first-order lagged droop controls and network losses. Specifically, we present theoretical results on the effects of filtering and the network topology on angle stability of droop-controlled microgrids. Towards this goal, we develop appropriate mathematical models for microgrid with low-pass-filtered droop control and thoroughly investigate the effects of the filtering on the dynamic stability of both lossy and lossless microgrids when subjected to small disturbances. We emphasize that the filter model used in this paper may represent an actual deployed low-pass filter used to reduce the measurement noise or may be a first-order approximation for a delay. 

A microgrid employs droop-control to modulate both the generated real power and reactive power in terms of measured frequency and voltages  respectively. Because our focus here is on small-signal stability, it is reasonable for us to focus solely on the modulation of the real power (i.e. $P-droop$). 
Specifically, just as in \cite{song2015small,song2017network,ainsworth2013structure,simpson2013synchronization}, we assume the bus voltage magnitude to be constant but not necessarily identical, and in consequence we can ignore the $Q-V$ dynamics. 
In analogy with the bulk grid, this assumption is often reasonable because there is a time-scale separation between the grid's voltage dynamics and angle dynamics: in particular, the typical time-frame for the angle-stability analysis ranges from 0.1 to 10 seconds, while for voltage stability it is in the range of 10-20 seconds. Under these assumptions, we can only evaluate the impacts of $P-droop$ on the small-signal stability of the microgrid. We acknowledge that there are circumstances where the voltage and angle dynamics for a distribution system may be entangled \cite{leitner2018small,yang2018small}. Such cases will require a full model that simultaneously considers the angle and voltage dynamics. However, the analysis becomes considerably more complicated and the simpler case considered here is important as a baseline even if there is some entanglement.

The major contributions are detailed below:



{\em 1) Microgrid Dynamic Models.} A nonlinear differential-algebraic equation (DAE) model is developed for lossless and lossy microgrids, with an accurate representation of droop control with filtering and heterogeneous load types. A structure-preserving linearized differential-equation model is also obtained to facilitate the small-signal stability analysis of the model.

{\em 2) Stability Analysis.}  We provide several graph-theoretic conditions for lossless and lossy microgrids that guarantees stability regardless of lags in droop control. For lossless network, we present sufficient and necessary conditions of stability and discuss graph-theoretic implications. For lossy networks without filtering, we present sufficient conditions for small-signal stability using the extension of the Sylvester's inertia theorem for non-symmetric matrices \cite{carlson1965rank} and derive equivalent concepts for critical lines for lossy case. When filtering are included in lossy case, we present a condition on filtering time constant when otherwise stable homogeneous micgrogrid becomes unstable. Finally, we discuss the special case of a radial lossy microgrid (as opposed to a mesh network), and verify that stability is maintained regardless of the filtering. 


{\em 3) Transient Response. }  The effects of filtering on the transient response of both lossless and lossy microgrid is shown using simulations carried on modified IEEE 9-bus and IEEE 57-bus test systems. It is observed that the stable lossless microgrid with droop-controllers having higher lags (due to higher filtering time constant) shows higher overshoot and settling time. An example simulation case study for lossy microgrid is also detailed that demonstrates the destabilization of otherwise stable equilibrium point on increasing the lag or filtering time constant. As a special case, for a radial lossy microgrid, we demonstrate that the small-signal stable equilibrium point remains stable regardless of the filtering in droop control similar to the lossless case. 

 It should be noted that the motivation for this work is closely related to communications aspects of the grid infrastructure. Specifically, a droop control in a microgrid requires filtering, and is subject to time-lag, because the controller is implemented using measurement data which must be transmitted to the inverter-based controls; this contrasts with droop controls in the bulk grid, which are implemented directly in the prime mover of synchronous generators. This characteristic in the microgrid yields a model with intrinsic lags or filtering, and hence motivates the analysis considered in the paper. From this perspective, our study is also aligned with a body of work at the interface of communications and controls engineering, which is concerned with feedback controls that are implemented over a communication channel (which may be subject to lags, communication errors, etc). 
\vspace{-0.2 cm}
\section{Modeling}
\label{sec:Modeling}
A mathematical model for the transient and small-signal dynamics of a droop-controlled microgrid operating in islanded mode is developed in this section. 
Models for the dynamics of a droop-controlled microgrid have been developed in several recent studies \cite{schiffer2013synchronization,ainsworth2013structure,simpson2013synchronization}. The model presented here incorporates: 1) time-lags and/or low-pass filtering that is present in real-power droop controls implemented at the inverters; and 2) heterogeneous buses including inverter based generator buses and frequency dependent/independent load buses and 3) losses in network lines.  First, a nonlinear DAE model is developed.  Then, two simplifications -- approximation of the algebraic equations as dynamic ones via singular perturbation, and linearization -- are undertaken to enable small-signal stability analysis.

\subsection{Microgrid Network: Nonlinear DAE Model Formulation}
\label{subsec:Microgrid Network: Nonlinear DAE Model Formulation}
Islanded operation of inverter-based microgrids requires an appropriate
mechanism for power-sharing among the DERs. One promising approach for power sharing is to regulate the power injection by each DER based on the local frequency measured at its connecting inverter, in a way that mimics droop control for synchronous generators in the bulk grid.
In an important recent work, Song and co-workers have modeled the transient dynamics of a microgrid network when such droop
controls are used, and then examined the small-signal stability of the network model via a linearization of the model \cite{song2017network}. The study \cite{song2017network} assumes an instantaneous feedback of the frequency signal by the droop controller, in analogy with the bulk grid.  However, in the microgrid setting, incorporating droop controls requires direct measurement of electrical frequency signals using power electronics, which thus necessitates use of low-pass filters for noise reduction and/or incurs time-lag. In this study, following \cite{song2015small}, we incorporate models of filtering/time-lag in droop controls into the microgrid-network model.  We also explicitly model frequency dependent heterogeneous load buses and line losses in the network.

We present the microgrid network model in three steps: 1) the model for the droop control (Section \ref{subsec:Droop-Controlled Inverter Model}); 2) the network power-flow and load model (Section \ref{subsec:Network Model}); and 3) the full DAE model which combines the network power-flow models with dynamic models of the DERs and loads (Section \ref{subsec:DAE Model for the Microgrid Network}). It is important to note that the microgrid network's dynamics involve both active and reactive processes; active and reactive droop controls affect angle and voltage magnitude, respectively \cite{guerrero2011hierarchical}. However, as mentioned earlier, based on the assumption that the angle dynamics are significantly faster than the voltage dynamics, we model only the real-power droop control ($P-droop$) and dynamics. Hence, only angle and frequency dynamics are considered. These assumptions are standard in the small-signal analysis of the distribution grid \cite{song2015small,ainsworth2013structure,simpson2013synchronization}.


\subsubsection{Droop-Controlled Inverter Model}
\label{subsec:Droop-Controlled Inverter Model}
\begin{figure}[t]
		\centering
		\includegraphics[width=3.2in]{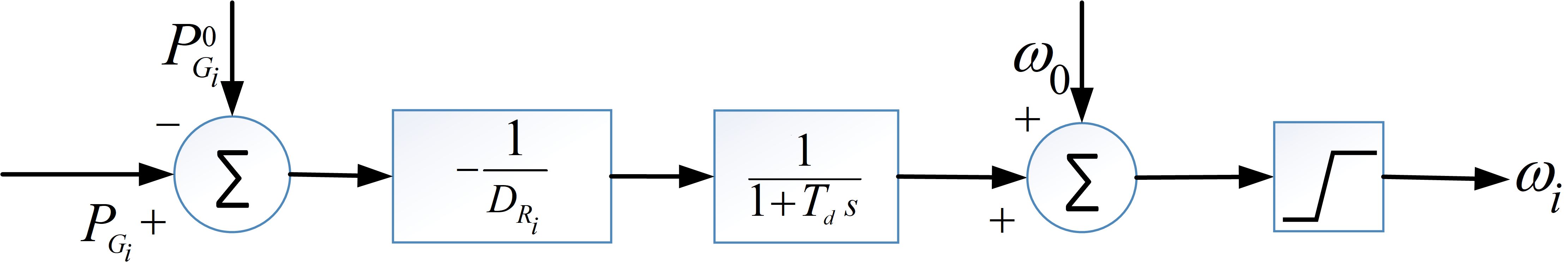}
		\vspace{-.3cm}
		\caption{Block Diagram of P-Droop Control with Filtering.}
		\label{fig:1}
\vspace{-0.5cm}		
	\end{figure}

The $P-droop$ control mechanism has an inherent delay, although it is typically small. In addition, a low pass filter is incorporated to suppress the high frequency variations in the measured power (see Fig. \ref{fig:1}). A few recent studies on small-signal stability of droop-controlled microgrids have represented such delay and/or filtering \cite{schiffer2013synchronization}; however, these studies have not fully incorporated the complexities of microgrids (e.g., losses, heterogeneity of loads). Here, we adopt a linear model for the droop control, which allows evaluation of the microgrid dynamics in the presence of these factors. Specifically, in our framework, the time delay and any filter in the $P-droop$ are abstracted as a first-order low pass filter. For an inverter at bus $i$, the power and frequency relationship is given in the Laplace domain as follows:
\vspace{-0.09cm}
\begin{equation}\label{eq1}
\frac{P^0_{G_i}-P_{G_i}(s)}{D_{R_i}} \times \frac{1}{1+T_{D_i}s}+\omega_ 0=\omega_{i}(s)
\vspace{-0.09cm}
\end{equation}
\noindent where, $\omega_0$ is the nominal angular frequency; $\omega_i(s)$ is the Laplace transform of frequency $\omega_i$;  $P^0_{G_i}$ is the reference power which is the generated active power at the nominal frequency; $P_{G_i}(s)$ is the Laplace transform of generated power $P_{G_i}$; $D_{R_i}\ge 0$ is the reciprocal of frequency droop gain of the inverter; and $T_{D_i}$ is the time constant of the low pass filter. Converting (\ref{eq1}) to the time domain and setting $\omega_0=0$ (since the droop control operates relative to the nominal rotational frequency e.g. $377$ rad/s in North America), we obtain:
	\begin{equation}\label{eq2}
	P_{G_i}\mathrm{=}P^0_{G_i}\mathrm{-}D_{R_i}\dot{{\theta }_i}\mathrm{-}D_{R_i}T_{D_i}\ddot{{\theta }_i}.
	\end{equation}
We notice that the droop control with filter involves a feedback of the angular acceleration, i.e. it mimics an inertial response.

\subsubsection{Network Model}
\label{subsec:Network Model}
A network model (power flow model) is developed here which incorporates lossy lines.  While stability analyses of the bulk grid often assume lossless transmission lines, microgrids operating at medium or low voltage level require modeling of line losses.

 We consider a connected microgrid network with $n$ buses, labeled $i=1,\hdots, n$. The admittance of the line between buses $i$ and $k$ obtained from the bus-admittance matrix is denoted by $Y_{ik}\angle \phi_{ik}= G_{ik}+jB_{ik}$, where $G_{ik} \leq 0$ and $B_{ik} \geq 0$ are respectively conductance and susceptance of the line according to the bus-admittance matrix. The power flow equation for bus $i$ is given as the following \cite{ainsworth2013structure}:
    \begin{eqnarray}\label{eq3}
	P_{G_i} &\mathrm{=}& P_{L_i}\mathrm{+}\sum_{k=1}^{n}{V_i V_k Y_{ij}~ \mathrm{cos} \mathrm{(}\theta_i\mathrm{-}\theta _k\mathrm{-} \phi_{ik}\mathrm{)} }  \\
    &\mathrm{=}& P_{L_i} + V_i^2 G_{ii}\mathrm{+}\sum_{k \in adj(i)}{V_i V_k Y_{ik}~\mathrm{cos} \mathrm{(}\theta_i\mathrm{-}\theta _k\mathrm{-} \phi_{ik}\mathrm{)}} \nonumber
\end{eqnarray}

\vspace{-0.1cm}
\noindent where, $P_{G_i}$ and $P_{L_i}$ are the active power generation and demand at bus $i$, respectively. $adj(i)$ refers to the set of buses that are adjacent to bus $i$ (or equivalently, the set of buses connected to bus $i$ via a distribution line). Note that the bus voltages are assumed to be constant but not necessarily identical at all the buses.
 
We use Bergen and Hill's model for frequency-dependent system loads \cite{bergen1981structure}. The load at bus $i$ can then be written as:
\begin{equation} \label{eq4}
P_{L_i}\mathrm{=}P^0_{L_i}\mathrm{+}D_{L_i}{\dot{\theta }}_i
\end{equation}
where, $P^0_{L_i}$ and $D_{L_i}\mathrm{\ge }\,\mathrm{ 0}$ are the nominal load and the frequency coefficient of the load at bus $i$.  In developing the DAE model for the dynamics, we distinguish frequency independent loads ($D_{L_i}=0$) from ones that have a frequency dependence.

\subsubsection{DAE Model for the Microgrid Network}
\label{subsec:DAE Model for the Microgrid Network}

The DAE model for the microgrid is detailed in this section. Let us first denote the set of buses that have inverters as ${\mathcal{V}}_A$. The governing differential equation for the bus $i\mathrm{\in }{\mathcal{V}}_A{\ }$ is derived using (\ref{eq2}-\ref{eq4}) and is given by:
 \begin{equation} \label{eq5}\begin{split}
D_{R_i}T_{D_i}\ddot{{\theta }_i}={}&(P^0_{G_i}\mathrm{-}P^0_{L_i}-V_i^2G_{ii}\mathrm{)-}\left(D_{L_i}\mathrm{+}D_{R_i}\right)\ {\dot{\theta }}_i\\&\mathrm{-}\sum_{k\mathrm{\ }\in \mathrm{\ }adj\mathrm{\ (}i\mathrm{)}}{V_iV_k{\mathrm{\ }Y}_{ik}~\mathrm{cos} \mathrm{(}\theta_i\mathrm{-}\theta _k\mathrm{-} \phi_{ik}\mathrm{)}},  i\mathrm{\in }{\mathcal{V}}_A
\end{split}
\end{equation}
We next present the governing equations for the buses that do not have inverter-based generators $(P_{G_i}\mathrm{=0})$. It is convenient to subdivide these buses into sets with and without frequency-dependent loads. We denote the set of buses that do not have inverters but have frequency dependent loads as ${\mathcal{V}}_B$. The governing equation for each  $i\mathrm{\in }{\mathcal{V}}_B$ is given by:
\vspace{-0.1cm}
\begin{equation} \label{eq6}\begin{split}
0\ \mathrm{=} (-P^0_{L_i}- & V_i^2G_{ii})\mathrm{-}D_{L_i}{\dot{\theta }}_i \\& \mathrm{-}\sum_{k\mathrm{\ }\in \mathrm{\ }adj\mathrm{\ (}i\mathrm{)}}{V_iV_k{\mathrm{\ }Y}_{ik}~\mathrm{cos} \mathrm{(}\theta_i\mathrm{-}\theta _k\mathrm{-} \phi_{ik}\mathrm{)}},  i\mathrm{\in}{\mathcal{V}}_B\mathrm{\ }
\end{split}
\end{equation}
\vspace{-0.1cm}
Next, we denote the set of buses that have neither inverter-based generation nor frequency dependent loads as ${\mathcal{V}}_C\mathrm{\ }$. The governing equation for each $i\mathrm{\in }{\mathcal{V}}_C$ is given by:
\begin{equation} \label{eq7}
0\ \mathrm{=}(-P^0_{L_i}-V_i^2G_{ii})-\sum_{k\mathrm{\ }\in \mathrm{\ }adj\mathrm{\ (}i\mathrm{)}}{V_iV_k{\mathrm{\ }Y}_{ik}~\mathrm{cos} \mathrm{(}\theta_i\mathrm{-}\theta _k\mathrm{-} \phi_{ik}\mathrm{)}},i\mathrm{\in }{\mathcal{V}}_C
\end{equation}


\vspace{-0.3cm}
\subsection{Simplified Models for Stability Analysis}
\label{subsec:Simplified Modeling for Stability Analysis}
We develop a linear differential-equation approximation of the nonlinear DAE to enable formal analysis of small-signal stability.  
First, a singular perturbation argument is applied to obtain a differential-equation approximation of the DAE in a way that maintains the topological structure of the network (hence, we refer to this as a structure-preserving model). Then, the nonlinear differential-equation model is linearized to obtain the small-signal model of the microgrid dynamics.

\subsubsection{Structure-Preserving Model}
\label{subsec:Structure-Preserving Model}
Stability analysis of DAE models is usually undertaken in one of the two ways: 1) the algebraic equations are solved (the so-called "Kron" reduction) and the stability of the resulting purely differential
equation is determined; or 2) a singular perturbation argument is applied to approximate the algebraic equations as fast dynamics yielding a differential-equation model, whereupon the stability can be assessed. The singular-perturbation approach has the advantage of preserving the topological structure of the microgrid, which is useful for developing topological characterizations of stability and other properties. Here, we develop a structure-preserving model approximation for the microgrid DAE model using the singular-perturbation approach.

Specifically, per a singular-perturbation approach, we approximate (\ref{eq6}) and (\ref{eq7}) as:

\vspace{-0.3cm}
\begin{equation} \label{eq8}\begin{split}
{\epsilon }_1\ddot{{\theta }_i}\ &\mathrm{=(0-}P^0_{L_i}-V_i^2G_{ii}\mathrm{)-}\left(\mathrm{0+}D_{L_i}\right)\ {\dot{\theta }}_i\\&
\mathrm{-}\sum_{k\mathrm{\ }\in \mathrm{\ }adj\mathrm{\ (}i\mathrm{)}}{V_iV_k{\mathrm{\ }Y}_{ik}\mathrm{cos} \mathrm{(}\theta_i\mathrm{-}\theta _k\mathrm{-} \phi_{ik}\mathrm{)}},  i\mathrm{\in }{\mathcal{V}}_B\mathrm{\ }
\end{split}
\end{equation}
\begin{equation} \label{eq9}\begin{split}
{\epsilon }_1\ddot{{\theta }_i}\ &\mathrm{=(0-}P^0_{L_i}-V_i^2G_{ii}\mathrm{)-}{\mathrm{\epsilon }}_{\mathrm{2}}{\dot{\theta }}_i\\&\mathrm{-}
\sum_{k\mathrm{\ }\in \mathrm{\ }adj\mathrm{\ (}i\mathrm{)}}{V_i V_k{\mathrm{\ }Y}_{ik}\mathrm{cos} \mathrm{(}\theta_i\mathrm{-}\theta _k\mathrm{-} \phi_{ik}\mathrm{)}},  i\mathrm{\in }{\mathcal{V}}_C\mathrm{\ }
\end{split}
\end{equation}
\noindent where, ${\epsilon }_{\mathrm{1}}$ and ${\epsilon }_{\mathrm{2}}$ are sufficiently small positive numbers.

According to the singular perturbation theory, the system described by (\ref{eq5}), (\ref{eq8}), (\ref{eq9}) and the system described by (\ref{eq5}), (\ref{eq6}), (\ref{eq7}) have the same equilibrium solutions. Additionally, provided that ${\epsilon }_{\mathrm{1}}$ and ${\epsilon }_{\mathrm{2}}$ are chosen appropriately (specifically, ${\epsilon }_{\mathrm{2}}$ is sufficiently small and ${\epsilon }_{\mathrm{1}}$ is on the order of ${\epsilon }_{\mathrm{2}}^2$), the local stability of the equilibrium is also maintained.  Hence, we use differential equation set (\ref{eq5}), (\ref{eq8}) and (\ref{eq9}) for our analysis. Here we clarify that because the non-linear model is a differential-algebraic equation, the constants ${\epsilon }_{\mathrm{1}}$ and ${\epsilon }_{\mathrm{2}}$ are exactly equal to zero in the correct linearization; the approximated equations are purely algebraic in the context of the swing-dynamics models. The use of the factors ${\epsilon }_{\mathrm{1}}$ and ${\epsilon }_{\mathrm{2}}$ is a construct which allows development of the formal results in the paper: the proofs give characterizations in the limit that these factors are zero. For the purpose of simulation, it is important that sufficiently small values for ${\epsilon }_{\mathrm{1}}$ and ${\epsilon }_{\mathrm{2}}$ are used (as example see Section \ref{sec:Lossless Case}). The appropriate bounds can be derived from the singular-perturbation literature \cite{kokotovic1999singular}. Another approach and the typical way of selecting these values is by validating simulations against numerical solutions of the differential-algebraic equations.

In close analogy with Song et al \cite{song2015small}, we find it convenient to present a compact form of these equations based on a graph-theoretic description of the microgrid. To do so, let us consider a directed graph $\mathcal{G}$ defined for the microgrid, where each bus is represented by a vertex, and the line between buses $i$ and $k$ corresponds to two directed edges $(i,k)$ and $(k,i)$. Hence, the vertex set $\mathcal{V}$ has cardinality $n$ and edge set $\mathcal{E}\subseteq \mathcal{V}\times\mathcal{V}$ has cardinality $2l$, where $l$ is the number of lines in the microgrid. It is convenient to arbitrarily label and order the edges, with edge $e_m$ ($m=1,\hdots, 2l$) representing edge $(i,k)$ of the graph.  Then, the incidence matrix $\boldsymbol{E} \in \mathbb{R}^{n\times 2l}$ of $\mathcal{G}$ is defined as  $E_{im}=1$ and $E_{km}=-1$ for each $e_m \in \mathcal{E}$, with all other entries being zero. Furthermore, the orientation matrix is defined for the directed graph (see \cite{goldin2013weight}) as a matrix $\boldsymbol{C}\in \mathbb{R}^{n\times 2l}$ with entries $C_{im}=1$ if $E_{im}=1$ and zero otherwise. Moreover, for the sake of simplicity, we denote a vector $\boldsymbol{x}=[x_1,x_2,...,x_n]^T \in \mathbb{R}^n$ as $\boldsymbol{x}=[x_i]$, and a diagonal matrix $\boldsymbol{D} \in \mathbb{R}^{n\times n}$ as $ diag(D_{11},D_{22}, \cdots ,D_{nn})$ where $D_{11},D_{22},\cdots ,D_{nn}$ are the diagonal entries. Using these notations, we can write (\ref{eq5}), (\ref{eq8}), and (\ref{eq9}) as a single vector equation, as follows:
\begin{equation} \label{eq10}
\boldsymbol{M}\ddot{\boldsymbol{\boldsymbol{\theta }}}\mathrm{=}\boldsymbol{P}\mathrm{-}\boldsymbol{\boldsymbol{D}}\dot{\boldsymbol{\theta}}-\boldsymbol{C} \boldsymbol{B}_l \cos{\boldsymbol{\mathrm{(}}{\boldsymbol{E}}^T \boldsymbol{\theta }-\boldsymbol{\phi})}
\end{equation}
where, $\boldsymbol{\theta } = [\theta_i] \in \mathbb{R}^n$ is the vector of bus phase angles; $\boldsymbol{M}~\mathrm{\in }\mathrm{\ }\mathbb{R}^{n\mathrm{\times }n}$ is a diagonal matrix imitating inertia with $M_{ii}\mathrm{=}D_{R_i}T_{D_i}$ if $i\mathrm{\in }{\mathcal{V}}_A\mathrm{\ }$, otherwise, $M_{ii}=\ {\epsilon }_1$;  $\boldsymbol{P}= [P_i] ~\mathrm{\in }\mathrm{\ }\mathbb{R}^n$ is a vector representing effective injected power in each bus with $\ P_i\mathrm{=}P^0_{G_i}\mathrm{-}P^0_{L_i}-V_i^2G_{ii}\mathrm{\ }$; $\boldsymbol{D}~\mathrm{\in }\mathrm{\ }\mathbb{R}^{n\mathrm{\times }n}$ is a diagonal matrix of total damping coefficients with $D_{ii}={\epsilon }_2$ if $i\mathrm{\in }{\mathcal{V}}_C$, otherwise, $\mathrm{\ }D_{ii}=\ D_{L_i}\mathrm{+}D_{R_i}$;  $\boldsymbol{\phi}=[\phi_{ik}] \in \mathbb{R}^{2l}$ is a vector of phase angles of the admittance of each directed lines  and $\boldsymbol{B}_l \in \mathbb{R}^{2l \times 2l}=diag(V_iV_k{\mathrm{\ }Y}_{ik})$ represents the magnitude of active power flow through the edges. Note that in (\ref{eq10}) the cosine function $cos(.)$ is meant in the hadamard sense (i.e. component-wise).

Equation (\ref{eq10}) is in suitable form to develop the small-signal model that is used to characterize the stability from a structural or graph-theoretic perspective.

\subsubsection{Small-Signal Model}
\label{subsec:Small-Signal Model}
Linearization is used to obtain a small-signal differential-equation model for the the microgrid network, which allows analysis of stability. For this purpose, let us define the state vector as  $\boldsymbol{x}{\mathrm{=}}\left(\genfrac{}{}{0pt}{}{\boldsymbol{\theta }}{\boldsymbol{\dot{\theta}}}\right)$. By linearizing (\ref{eq10}) around the equilibrium point ${\boldsymbol{x}}^{{\mathrm{0}}}{\mathrm{=}}\left(\genfrac{}{}{0pt}{}{{\boldsymbol{\theta }}^{\mathrm{0}}}{{\boldsymbol{\dot{\theta} }}^{{\mathrm{0}}}}\right)$, we get
\vspace{-0.2cm}
\begin{equation} \label{eq11}
\boldsymbol{\Delta }\dot{\boldsymbol{x}}=\left(\genfrac{}{}{0pt}{}{\boldsymbol{\mathrm{\Delta }}\dot{\boldsymbol{\theta }}}{\boldsymbol{\mathrm{\Delta }}\ddot{\boldsymbol{\theta }}}\right)=\boldsymbol{J}(\boldsymbol{x}^0)\left(\genfrac{}{}{0pt}{}{\boldsymbol{\mathrm{\Delta }}\boldsymbol{\theta }}{\boldsymbol{\mathrm{\Delta }}\dot{\boldsymbol{\theta}}}\right)=\boldsymbol{J}(\boldsymbol{x}^0)\boldsymbol{\mathrm{\Delta }}\boldsymbol{x}
\end{equation}
 where $\boldsymbol{J}\left({\boldsymbol{x}}^{\boldsymbol{\mathrm{0}}}\right)$ is the Jacobian matrix of the nonlinear model evaluated at the equilibrium.  The Jacobian is found to be:
 \vspace{-0.1cm}
\begin{equation} \label{eq12}
\boldsymbol{J}(\boldsymbol{x}^0){\ }\boldsymbol{\mathrm{=\ }}\left[ \begin{array}{cc}
{\boldsymbol{\mathrm{0}}}_{{n\times n}} & {\boldsymbol{I}}_{{n\times n}} \\
-{\boldsymbol{M}}^{-1}\boldsymbol{\boldsymbol{C}}\boldsymbol{\mathrm{\ }}\boldsymbol{W}(\boldsymbol{\theta}^0){\boldsymbol{E}}^{{T}} & \boldsymbol{\mathrm{-}}{\boldsymbol{M}}^{{\mathrm{-}}{\mathrm{1}}}\boldsymbol{D} \end{array}
\right]
\end{equation}
where, $\boldsymbol{W}(\boldsymbol{\theta})=\frac{\partial  \boldsymbol{B}_l \cos {\boldsymbol{\mathrm{(}}{\boldsymbol{E}}^T \boldsymbol{\theta }-\boldsymbol{\phi})}}{\partial ({\boldsymbol{E}}^T \boldsymbol{\theta }-\boldsymbol{\phi})}= diag \big(-\boldsymbol{B}_l \sin {\boldsymbol{\mathrm{(}}{\boldsymbol{E}}^T \boldsymbol{\theta }-\boldsymbol{\phi}) \big) }$ $\in \mathbb{R}^{2l \times 2l}$. Here, $\boldsymbol{0}_{n \times n}$ and $\boldsymbol{I}_{n \times n}$ denote $n \times n$ zero matrix and $n \times n$ unit matrix respectively. Note that as (\ref{eq10}) depends on the relative angles of the buses, therefore, $\boldsymbol{x}^0$ and $\boldsymbol{x}^0 +k \boldsymbol{v}_0$ refer to physically same equilibrium points of (\ref{eq10}) where, ${\boldsymbol{v}_0=\begin{bmatrix} \boldsymbol{1}_n\\ {\bf 0}_n \end{bmatrix}}$, and $\boldsymbol{1}_n$ and $\boldsymbol{0 }_n$ are vectors in $\mathbb{R}^n$ with all entries equal to $1$ and $0$, respectively. Therefore, the linearization undertaken in (\ref{eq10}) can be understood as a linearization over the manifold $\boldsymbol{x}^0 +k \boldsymbol{v}_0$ and thus, there exists an invariant manifold where all the angles are synchronized and all the frequency deviations are zero in the linearized system (\ref{eq11}). Again in analogy with \cite{song2015small}, the form of (\ref{eq12}) motivates us to define $\boldsymbol{W}(\boldsymbol{\theta}^0)$ as the edge weights of the graph $\mathcal{G}$. This definition allows us to rewrite the dynamics in terms of the (directed) Laplacian matrix of the graph, which is helpful for the analysis of the stability.  Like \cite{song2015small}, we refer to  the weighted directed graph $\mathcal{G}(\boldsymbol{x}^0)=(\mathcal{V} ,\mathcal{E},\boldsymbol{W}(\boldsymbol{\theta}^0))$ as \textit{active power flow graph}. We also note that the edge weights in the graph $\mathcal{G}(\boldsymbol{x}^0)$ depend on the operating point $\boldsymbol{x}^0$. The Laplacian matrix for this directed weighted graph can be written as $\boldsymbol{L}(\mathcal{G}(\boldsymbol{x}^0))=\boldsymbol{C}\boldsymbol{\ }\boldsymbol{W} (\boldsymbol{\theta}^0) {\boldsymbol{E}}^{{T}}$. Therefore, from (\ref{eq11}) and  (\ref{eq12}) we can write:
\begin{equation} \label{eq13}
\boldsymbol{\Delta }\dot{\boldsymbol{x}}= \boldsymbol{J}(\boldsymbol{x}^0) \boldsymbol{\mathrm{\Delta }}\boldsymbol{x} = \left[ \begin{array}{cc}
{\boldsymbol{\mathrm{0}}}_{{n\times n}} & {\boldsymbol{I}}_{{n\times n}} \\
-{\boldsymbol{M}}^{-1}\boldsymbol{L}(\mathcal{G}(\boldsymbol{x}^0)) & \boldsymbol{\mathrm{-}}{\boldsymbol{M}}^{{\mathrm{-}}{\mathrm{1}}}\boldsymbol{D} \end{array}\right] \boldsymbol{\mathrm{\Delta }}\boldsymbol{x}
\end{equation} 

The developed small-signal model can be further simplified if we assume the microgrid network is lossless, or there is no filtering in the droop control. For a lossless microgrid, the admittance of the line between buses $i$ and $k$ is given by $Y_{ik}\angle90^{\circ} =jB_{ik}$ and therefore, the edge weights of both $(i,k)$ and $(k,i)$ edges become same and equal to $V_i V_k B_{ik}\cos{(\theta_i -\theta_k)}$ in active power flow graph $\mathcal{G}$. Thus, $\mathcal{G}$ can be considered as an undirected graph where each line between bus $i$ and $k$ corresponds to an undirected edge $(i,k)$. For undirected graph, the Laplacian matrix is symmetric and can be written as $\boldsymbol{L}(\mathcal{G}(\boldsymbol{x}^0))=\boldsymbol{E}_u \boldsymbol{W}_u(\boldsymbol{\theta}^0) \boldsymbol{E}_u^{T}$ where, $\boldsymbol{E}_u \in \mathbb{R}^{n\times l}$ is the incidence matrix and $\boldsymbol{W}_u(\boldsymbol{\theta}^0)= diag(V_iV_k{\mathrm{\ }Y}_{ik}) \cos {(\boldsymbol{E}_u^T \boldsymbol{\theta}^0)}=  \boldsymbol{B}_{l,u} \cos {(\boldsymbol{E}_u^T \boldsymbol{\theta}^0)} \in \mathbb{R}^{l \times l}$ is the weights of the edges in the undirected graph $\mathcal{G}$. In the case when there is no filtering in any of the droop control, the structure preserving model (\ref{eq10}) reduces to $\boldsymbol{\boldsymbol{D}}\dot{\boldsymbol{\theta}}= \boldsymbol{P}-\boldsymbol{C} \boldsymbol{B}_l \cos{(\boldsymbol{E}^T \boldsymbol{\theta }-\boldsymbol{\phi})}$ and small-signal model (\ref{eq13}) reduces to the following:
\begin{equation} \label{eq14}
\boldsymbol{\Delta }\dot{\boldsymbol{x}}= \boldsymbol{J}(\boldsymbol{x}^0) \boldsymbol{\mathrm{\Delta }}\boldsymbol{x} = -\boldsymbol{D}^{-1} \boldsymbol{L}(\mathcal{G}(\boldsymbol{x}^0)) ~\boldsymbol{\mathrm{\Delta }}\boldsymbol{x} 
\end{equation} 
where, the state vector $\boldsymbol{x}= \boldsymbol{\theta}$. Note that (\ref{eq14}) is similar to the model developed in \cite{song2015small} but here $\boldsymbol{L}(\mathcal{G}(\boldsymbol{x}^0))$ can be asymmetric also. 

\section{Small-Signal stability analysis}
\label{sec:Small disturbance stability analysis}

Here, we analyze the small-signal stability of the microgrid, as defined by the small-signal model (\ref{eq13}). Our main aim is to understand the impact of the lag/ filtering time constant and the network topology on stability.  In our analysis, we distinguish stability characteristics for lossless vs. lossy networks and also for general vs. radial topologies. 

Before proceeding, first we review the notion of small-signal stability in our context, see the standard literature on power-system analysis \cite{kundur1994power}. To begin, we note that the vector ${\boldsymbol{v}_0=\begin{bmatrix} \boldsymbol{1}_n \\ {\bf 0}_n \end{bmatrix}}$ is in the null space of the Jacobian matrix $\boldsymbol{J}(\boldsymbol{x}^0)$. Therefore, as mentioned earlier, the manifold where all the angles are synchronized and all the frequency deviations are zero is an invariant manifold. The existence of an invariant manifold, rather than a single invariant point, is apparent as the non-linear dynamics (\ref{eq10}) depends on the angle differences and frequencies only. Therefore, the small-signal model (\ref{eq13}) is said to be  stable when the invariant manifold where all the angles are synchronized and all the frequency deviations are zero is asymptotically stable in the sense of Lyapunov \cite{nonlinear}. With this context, next we present our results on the conditions for stability of lossless microgrid, and then we expand the results for the general lossy microgrid cases.

\subsection{Small-Signal Stability Analysis of Lossless Microgrid}
\label{sec:Small-Signal Stability Analysis of Lossless Microgrid}

There has been a considerable effort to characterize microgrid stability for a lossless network model \cite{ainsworth2013structure,simpson2013synchronization,song2015small}. While we focus on the lossy case here, in this subsection we review and develop the small-signal stability analysis of microgrids when lossless assumption is made, for the sake of completeness. Recall that for lossless case the active power flow graph is undirected and the Laplacian matrix is symmetric. 


We begin by reviewing the Laplacian-based necessary and sufficient condition for stability for the lossless microgrid when there is no filtering/lag in the droop control. The result is similar to that of Lemma 1 and the Theorem 1 of  \cite{song2015small} and therefore presented without proof.

\medskip

\noindent \textbf{Theorem 1:}  \textit{ For a lossless microgrid without lag (filtering) in the P-droop control, the small-signal model (\ref{eq14}) is asymptotically stable if and only if the Laplacian  $\boldsymbol{L}(\mathcal{G}(\boldsymbol{x}^0))$ of the active power flow graph is positive semidefinite with a non-repeated eigenvalue at zero.}

\medskip

Next, the stability of the microgrid is examined when the droop control is subject to filtering. Note that the small-signal model described by (\ref{eq13}) is similar to the structure-preserving classical small-signal model developed for the bulk power grid with synchronous generators. The droop-controlled network with filtering has been considered in \cite{schiffer2013synchronization} with the assumption of uniform damping, however the analysis does not give necessary and sufficient conditions, and also does not consider the singular perturbation model which captures heterogeneous load. To develop a necessary and sufficient condition, we instead exploit a result that was developed in \cite{song2017network} for stability analysis of the bulk grid. Specifically, our result uses  Theorem 1 of \cite{song2017network}, which is based on relating the number of eigenvalues of the Jacobian having positive real part with the number of eigenvalues of the Laplacian having negative real part using the Sylvester's law of inertia \cite{horn1990matrix} and the Proposition 4-2 of \cite{chiang1989closest}. Here, we express the small-signal model for the droop-controlled lossless microgrid network with filtering in an appropriate form, and then directly apply Theorem 1 of \cite{song2017network}. Here is the criterion:

\medskip

\noindent \textbf{Theorem 2:}  \textit{Assume that the zero eigenvalue of the Laplacian matrix   $\boldsymbol{L}(\mathcal{G}(\boldsymbol{x}^0))$ of the active power-flow graph has algebraic multiplicity of $1$ (i.e., is non-repeated).  The small-signal model of a lossless microgrid as described by (\ref{eq13}) is asymptotically stable for any filtering time constant if and only if the Laplacian matrix is positive semi-definite.  If the condition is not satisfied, the small-signal model is not stable for any filtering time constant.}
\medskip

\noindent \textbf{Proof:}

We show that the state matrix of the linearized model (\ref{eq13}) has the same form as that of the Jacobian matrix in \cite{song2017network} and, therefore, Theorem 1 of \cite{song2017network} establishes our result. Since the nonlinear model described by (\ref{eq10}) and its small-signal model (\ref{eq13}) depends on the relative angle of the bus voltages, without loss of generality, we take the voltage angle at bus $n$ as the reference and apply the following transformation, $\boldsymbol{\alpha}=\boldsymbol{T \theta}$, where, $\boldsymbol{\alpha} \in \mathbb{R}^{n-1}$ is the vector of relative bus voltage angles with respect to bus $n$ and $\boldsymbol{T}= [\boldsymbol{I}_{(n-1) \times (n-1)}$ $-\boldsymbol{1}_{n-1} ] \in \mathbb{R}^{(n-1) \times n}$.  
Let us define the new state vector as $\boldsymbol{x}_n= \left(\genfrac{}{}{0pt}{}{\boldsymbol{\alpha }}{\boldsymbol{\dot{\theta}}}\right)$. The linearization around the equilibrium point ${\boldsymbol{x}}_n^{{\mathrm{0}}}=\left(\genfrac{}{}{0pt}{}{{\boldsymbol{\alpha }}^{\mathrm{0}}}{{\boldsymbol{\dot{\theta} }}^{{\mathrm{0}}}}\right)$ yields $\boldsymbol{\Delta }\dot{\boldsymbol{x}}_n=\boldsymbol{J}(\boldsymbol{x}_n^0)\boldsymbol{\mathrm{\Delta }}\boldsymbol{x}_n$. Using (\ref{eq13}), where $\boldsymbol{L}(\mathcal{G}(\boldsymbol{x}^0))=\boldsymbol{E}_u \boldsymbol{W}_u(\boldsymbol{\theta}^0) \boldsymbol{E}_u^{T}$ for the undirected graph, we obtain $\boldsymbol{J}(\boldsymbol{x}_n^0)$ as:
\begin{eqnarray} \label{eq15}
\boldsymbol{J}(\boldsymbol{x}_n^0) &=& diag (\boldsymbol{T},\boldsymbol{I}_{n\times n}) ~\boldsymbol{J}(\boldsymbol{x}^0)~diag (\boldsymbol{T}^{\dagger},\boldsymbol{I}_{n\times n}) \nonumber \\
&=& \left[ \begin{array}{cc}
\boldsymbol{0}_{(n-1)\times (n-1)} & \boldsymbol{T}\\
-\boldsymbol{M}^{-1}\boldsymbol{T}^T\boldsymbol{E}_s\boldsymbol{W}_u(\boldsymbol{\theta}^0)\boldsymbol{E}_s^T & -\boldsymbol{M}^{-1}\boldsymbol{D} \end{array}
\right] \nonumber \\
\end{eqnarray}
Here, we have used $\boldsymbol{E}_u= \boldsymbol{T}^T \boldsymbol{E}_s$, where $\boldsymbol{E}_s \in \mathbb{R}^{(n-1) \times l}$ consists of the first $(n-1)$ rows of $\boldsymbol{E}_u$ and $\boldsymbol{T}^{\dagger}= \boldsymbol{T}^T (\boldsymbol{T}\boldsymbol{T}^T)^{-1}$, where $\boldsymbol{T}^{\dagger}$ is the right pseudo-inverse of $\boldsymbol{T}$. Note that (\ref{eq15}) is identical to the equation (7) of \cite{song2017network} with $\boldsymbol{T}_L=0,~\boldsymbol{T}_G=\boldsymbol{T},~\boldsymbol{D}_G= \boldsymbol{D}$. Therefore, the necessary and sufficient condition of asymptotic stability given in Theorem 1 of \cite{song2017network} implies that the manifold where all of the angles are synchronized and all the frequency deviations are zero is asymptotically stable, if and only if, the Laplacian matrix $\boldsymbol{L}(\mathcal{G}(\boldsymbol{x}^0))$ is positive semi-definite. Note that from (13) of \cite{song2017network}, it is clear that the assumption made on the nonsingularity of $\boldsymbol{F}(\boldsymbol{\alpha}^0)=\boldsymbol{E}_s \boldsymbol{W}_u(\boldsymbol{\theta}^0) \boldsymbol{E}_s^{T}$ in \cite{song2017network} is equivalent to the Laplacian $\boldsymbol{L}(\mathcal{G}(\boldsymbol{x}^0))$ having a non-repeated eigenvalue at zero.\hfill\(\Box\)

Theorem 2 demonstrates that filtering in the real-power droop control does not alter the microgrid's small-signal stability, provided that the model is lossless. Algebraic and graph-theoretic conditions under which $\boldsymbol{L}(\mathcal{G}(\boldsymbol{x}^0))$ satisfies the conditions --  i.e., the matrix is positive semidefinite, and has a non-repeated eigenvalue at zero -- can readily be determined. The Laplacian is guaranteed to be positive semi-definite and has a non-repeated eigenvalue at zero, if its off-diagonal entries are nonnegative and the matrix is irreducible (equivalently, its associated graph is connected and has positive edge weights). For the Laplacian matrix in our formulation, the weight of the edge between vertices $i$ and $k$ is given by  $V_i V_k B_{ik}\cos{(\theta_i -\theta_k)}$.  This weight is positive if the difference between the power-flow bus angle across the corresponding line is less than $\pi/2$, or more generally $|\theta_i^0-\theta_k^0|~ mod ~2\pi < \pi/2$. If this condition is not satisfied by any line, the line is called {\em critical} \cite{song2015small,song2017network}. As a special case, if the distribution grid is {\em{radial (i.e. the graph of the Laplacian is a tree)}}, there exists a solution to the power-flow equation for which no line is critical (see also the discussion for the lossy case in the next section).  Hence this equilibrium (manifold) is asymptotically stable for any lag. These observations are summarized in the following corollaries to Theorem 2 (presented without proof, since they are automatic consequences of the graph-theoretic analysis of the Laplacian).

\medskip

\noindent \textbf{Corollary 1:}  \textit{Consider the active power flow graph $\mathcal{G}(\boldsymbol{x}^0)$ at the equilibrium of interest.  If all the edge weights are positive or equivalently, there are no critical lines, then the small-signal model (\ref{eq13}) of a lossless microgrid  is asymptotically stable for any filtering time constant.}
\medskip

\noindent \textbf{Corollary 2:}  \textit{Consider a microgrid whose active power flow graph is radial. The power flow has an equilibrium solution for which no line is critical.  For this equilibrium, the small-signal model is asymptotically stable for any filtering time constant.}
\medskip

\noindent{\em{Discussions:}} Note that the stability criterion given in Theorem 2 is independent of the droop control's lag (time constant of the low-pass filter), hence it implies Theorem 1 as a special case. The condition given in Corollary 1 is a sufficient condition: stability may be achieved even when some lines are critical. We refer the reader to  \cite{song2015small,song2017network} for characterization of such cases. Even though any lags do not influence stability under broad conditions in the lossless case, simulations suggest that lags may have significant impact on the damping and settling time of the microgrid, see Section \ref{sec:simulation}.
\vspace{-0.1cm}
\subsection{Small-Signal Stability Analysis of Lossy Microgrid}
\label{sec:Small disturbance stability analysis of lossy microgrid}
Unlike the bulk grid, the line resistances for MV/LV power distribution systems are relatively high. Thus, the assumption that microgrid is lossless is often inaccurate. In this subsection, we analyze small-signal stability  for the general microgrid model.  
For convenience of presentation, we refer to the model as the lossy microgrid model, although it encompasses the lossless case. We stress that the challenge in analyzing the lossy case comes from the fact that the the Laplacian matrix associated with the power-flow graph is non-symmetric.

 To develop results for the general lossy case, we use an extension of the Sylvester's inertia theorem for non-symmetric matrices. First, we define mathematical notions on inertia\footnote{We note that this term ``inertia" is unrelated to the notion of generator inertia; we use this language in keeping with the terminology used in \cite{horn1990matrix}.} \cite{horn1990matrix}. For a square matrix $\boldsymbol{A}$,  we denote the number of eigenvalues of $\boldsymbol{A}$ with positive, negative and zero real parts as $i_+(\boldsymbol{A}), i_-(\boldsymbol{A})$ and $i_0(\boldsymbol{A})$, respectively. We call the ordered triple, $\big(i_+(\boldsymbol{A}), i_-(\boldsymbol{A}), i_0(\boldsymbol{A})\big)$, the inertia of $\boldsymbol{A}$. We use the notation $sym(\boldsymbol{A})$ to indicate the symmetric part of $\boldsymbol{A}$ which is given by $\frac {(\boldsymbol{A}+ \boldsymbol{A}^T)} {2}$. Next, we state an extension of Sylvester's inertia theorem, which is contained in Theorem 3 of \cite{carlson1965rank}.

\medskip

\noindent \textbf{Lemma 1:}  \textit{ Let $\boldsymbol{A}$ be a square matrix and $\boldsymbol{H}$ be a symmetric matrix of same dimension. If $sym(\boldsymbol{AH})$ is positive semidefinite with rank $r$, then $\big(r-i_+(\boldsymbol{A})\big) \leq i_-(\boldsymbol{H})$ \cite{carlson1965rank}.}
\medskip

This lemma allows us to develop Laplacian-based sufficient condition for stability for a lossy microgrid without any filtering in the droop controls. 

\medskip

\noindent \textbf{Theorem 3:}  \textit{Consider a lossy microgrid without lag (filtering) in the P-droop control, and assume that the symmetric part of the Laplacian matrix of the active power flow graph $sym(\boldsymbol{L}(\mathcal{G}(\boldsymbol{x}^0)))$ is positive semidefinite and has a non-repeated eigenvalue at zero. Then the small-signal model described by (\ref{eq14}) is asymptotically stable.}

\medskip

\noindent \textbf{Proof:}

From (\ref{eq14}), it follows that: $\boldsymbol{L}(\mathcal{G}(\boldsymbol{x}^0)) = (-\boldsymbol{D}\boldsymbol{J}(\boldsymbol{x}^0) \boldsymbol{D}^{-1}) \boldsymbol{D}$. When $sym(\boldsymbol{L}(\mathcal{G}(\boldsymbol{x}^0)))$ is positive semidefinite and has a non-repeated eigenvalue at 0, the rank $r$ of $sym(\boldsymbol{L}(\mathcal{G}(\boldsymbol{x}^0)))$ is $(n-1)$. Now we use Lemma 1 considering $\boldsymbol{A} \equiv (-\boldsymbol{D}\boldsymbol{J}(\boldsymbol{x}^0) \boldsymbol{D}^{-1})$ and $\boldsymbol{H} \equiv \boldsymbol{D}$. According to Lemma 1 we can say: $i_-(\boldsymbol{J}(\boldsymbol{x}^0))=i_+(-\boldsymbol{D}\boldsymbol{J}(\boldsymbol{x}^0) \boldsymbol{D}^{-1})\geq (n-1)$ since  $i_-(\boldsymbol{D})= 0$. Since, $\boldsymbol{1}_n$ is a vector in the null space of $(\boldsymbol{J}(\boldsymbol{x}^0))$, the inertia of  $(\boldsymbol{J}(\boldsymbol{x}^0))$ is given by $(0,n-1,1).$ Hence, the small-signal model described by (\ref{eq14}) is asymptotically stable in our context.
\hfill\(\Box\)

Unlike the lossless microgrid, the Laplacian based condition given in Theorem 3 is not a necessary condition. We can develop another sufficient condition for the small-signal stability of lossy micorgrid without filtering in the droop control, in terms of the edge weights of the active power flow graph.\vspace{0.1cm} 

\noindent \textbf{Lemma 2:}  \textit{Consider a lossy microgrid without lag (filtering) in the P-droop control. If all the edge weights in $\boldsymbol{W}(\boldsymbol{\theta})$ are positive, the small-signal model described by (\ref{eq14}) is asymptotically stable.}

\noindent \textbf{Proof:}

If the edge weights are all positive, then the Laplacian $\boldsymbol{L}(\mathcal{G}(\boldsymbol{x}^0))$ is the negative of a Metzler matrix, which further has the property that each row sums to zero. Since, $\boldsymbol{D}$ is diagonal and positive definite, $\boldsymbol{D}^{-1} \boldsymbol{L}(\mathcal{G}(\boldsymbol{x}^0))$ is also the negative of a Metzler matrix, for which each row sums to zero. From Gershgorin's circle theorem  \cite{horn1990matrix} and the connectivity assumption of the microgrid, it is straightforward to verify that that the small-signal model (\ref{eq14}) is asymptotically stable.
\hfill\(\Box\)

Lemma 2 motivates us to extend the notion of critical lines for the lossy network. For the lossy case,  the edges $(i,k)$ and $(k,i)$ will both have positive weights if $0 < (\phi_{ik}\pm |\theta_i^0-\theta _k^0|)~ \mbox{mod}~ 2\pi < \pi$ is satisfied. We define a line to be  critical when the above angle condition is not satisfied. For a directed graph, positive edge weights do not guarantee that the symmetric part of the Laplacian is positive semidefinite. Therefore, the sufficient condition mentioned in Theorem 3 may not be necessary. As a counterexample, if $\boldsymbol{L}(\mathcal{G}(\boldsymbol{x}^0))=[0.2, -0.1, -0.1; -0.6, 0.7, -0.1; -0.6, -0.1, 0.7]$ then all the edge weights are positive and therefore, by Lemma 2,  small-signal model (\ref{eq14}) is stable. However, $sym(\boldsymbol{L}(\mathcal{G}(\boldsymbol{x}^0)))$ is not positive semidefinite, as one of its eigenvalue is -0.1339.

We now focus our attention to the case when the droop-controls in the lossy microgrid have filtering. It can be shown that a Laplacian-based sufficient condition similar to Theorem 3 does not hold anymore, when there is filtering in the active power droop control. To show this, we first study a homogeneous microgrid where each bus has inverter with the same frequency droop gain and filtering time constant. For this special model, we derive a condition on filtering time constant for which (\ref{eq13}) becomes unstable.\vspace{0.1cm}

\noindent \textbf{Lemma 3:}  \textit{Consider a homogeneous lossy microgrid where each bus has inverter with same droop control. Let the reciprocal of frequency droop gain and the time constant of low pass filter in the droop control of each bus are $d$ and $r$ respectively. If $r> \frac{\mbox{Re}(\mu)}{(\mbox{Im}(\mu))^2} d$, where $\mu$ is any of the eigenvalues of the Laplacian
$\boldsymbol{L}(\mathcal{G}(\boldsymbol{x}^0))$ of the active power flow graph and $\mbox{Re}(\mu)$ and $\mbox{Im}(\mu)$ are its real and imaginary part respectively, the small-signal model (\ref{eq13}) is unstable.}

\noindent \textbf{Proof:}
Here, $\boldsymbol{M}=rd \boldsymbol{I}_{n \times n}$ and $\boldsymbol{D}=d \boldsymbol{I}_{n \times n}$.  First we will simplify the characteristic equation of the Jacobian $\boldsymbol{J}(\boldsymbol{x}^0)$ utilizing the properties of the determinant of block matrix. According to (13) the characteristic equation of $\boldsymbol{J}(\boldsymbol{x}^0)$ can be written as follows: ${\mathrm{det} \left(\lambda \boldsymbol{I}_{2n \times 2n}-\boldsymbol{J}(\boldsymbol{x}^0)\right)}={\mathrm{det} \left(\lambda \left(\lambda +\frac{1}{r}\right)\boldsymbol{I}_{n \times n}+\frac{1}{rd} \boldsymbol{L}(\mathcal{G}(\boldsymbol{x}^0))\right)}$. If $\boldsymbol{L}(\mathcal{G}(\boldsymbol{x}^0))$ has an eigenvalue $\mu $, then by comparing between the characteristic equations of $\boldsymbol{L}(\mathcal{G}(\boldsymbol{x}^0))$ and $\boldsymbol{J}(\boldsymbol{x}^0)$ we get: $\lambda \left(\lambda +\frac{1}{r}\right)=\frac{-\mu }{rd}$. Equivalently, $rd\ {\lambda }^2+d\ \lambda +\mu =0$. Now if $r> \frac{\mbox{Re}(\mu)}{(\mbox{Im}(\mu))^2} d$ for any eigenvalue $\mu$ of  $\boldsymbol{L}(\mathcal{G}(\boldsymbol{x}^0))$, then the system has eigenvalue in right half plane, making the lossy network unstable.
\hfill\(\Box\)

Lemma 3 shows that increasing the filtering time constant (lowering the bandwidth of the filters) in droop controls may destabilize the small-signal model. In particular, for the lossy case, $\boldsymbol{L}(\mathcal{G}(\boldsymbol{x}^0))$ is non-symmetric and can potentially admit complex eigenvalues, and hence stability may be lost for sufficiently small filter bandwidth. For instance, the non-symmetric Laplacian matrix 
$\boldsymbol{L}(\mathcal{G}(\boldsymbol{x}^0))= [0.9,$ $-0.8,-0.1;-0.1,0.9,-0.8;-0.8,-0.1,0.9]$ has positive edge weights and also has positive semidefinite $sym(\boldsymbol{L}(\mathcal{G}(\boldsymbol{x}^0)))$ with a non-repeated eigenvalue at zero, hence stability of the microgrid model is guaranteed in the non-lagged case.  The eigenvalues of the Laplacian for this example are $\mu =0, 1.3500 \pm j 0.6062$, i.e. the Laplacian has a pair of complex eigenvalues. Since $\boldsymbol{L}(\mathcal{G}(\boldsymbol{x}^0))$ has right-half-plane but complex eigenvalues, small filtering time constant will maintain stability while large ones will cause instability per Lemma 3. In Section \ref{sec:simulation}, we construct a small 3-bus network that exhibits instability for larger time constant. From the expression of $\lambda$, it is also evident that within the stability region, the increase of time constant decreases the damping and increases the settling time. Our simulation results shown in  Section \ref{sec:simulation} also illustrate this transient behavior. However, if the Laplacian matrix $\boldsymbol{L}(\mathcal{G}(\boldsymbol{x}^0))$ satisfies certain special properties, it can be shown that the small-signal model (\ref{eq13}) is asymptotically stable for any lag. For example, stability can be guaranteed if the Laplacian matrix is similar to a positive semidefinite matrix, via a diagonal similarity transform.  We present the result next:\vspace{0.1cm}

\noindent \textbf{Theorem 4:}  \textit{Consider a lossy microgrid for which the Laplacian $\boldsymbol{L}(\mathcal{G}(\boldsymbol{x}^0))$ of the active power flow graph has a non-repeated eigenvalue at zero, and further there exists a diagonal matrix $\boldsymbol{K}$ such that $\boldsymbol{L}_s(\mathcal{G}(\boldsymbol{x}^0))= \boldsymbol{K}\boldsymbol{L}(\mathcal{G}(\boldsymbol{x}^0)) \boldsymbol{K}^{-1}$ is positive semidefinite.  Then the small-signal model (\ref{eq13}) of the lossy microgrid is asymptotically stable for any filtering time constant}\vspace{0.1cm}.


\noindent \textbf{Proof:}

Consider the following similarity transformation of the linearized model's state matrix:
\begin{eqnarray} \label{eq16}\vspace{-0.3cm}
\boldsymbol{J}_s(\boldsymbol{x}^0) &=& diag(\boldsymbol{K},\boldsymbol{K})~ \boldsymbol{J}(\boldsymbol{x}^0) ~ diag (\boldsymbol{K}^{-1},\boldsymbol{K}^{-1}) \nonumber \\
&=& \left[ \begin{array}{cc}
\boldsymbol{0}_{n\times n} & \boldsymbol{I}_{n\times n} \\
-\boldsymbol{M}^{-1}\boldsymbol{L}_s(\mathcal{G}(\boldsymbol{x}^0)) & -\boldsymbol{M}^{-1}\boldsymbol{D} \end{array}
\right]\vspace{-0.1cm}
\end{eqnarray}

Note, in (\ref{eq16}) we have used the fact that the matrix multiplication of diagonal matrices of same order is commutative. Now since the similarity transformation does not change the eigenvalues and $\boldsymbol{L}_s(\mathcal{G}(\boldsymbol{x}^0))$ is positive semidefinite with a non-repeated eigenvalue at zero, it is immediate from Theorem 2 that the small-signal models governed by $\boldsymbol{J}_s(\boldsymbol{x}^0)$ and hence $\boldsymbol{J}(\boldsymbol{x}^0)$ are asymptotically stable.
\hfill\(\Box\)

Next we identify graph theoretic conditions for which the Laplacian $\boldsymbol{L}(\mathcal{G}(\boldsymbol{x}^0))$ has a non-repeated eigenvalue at zero and there exists a diagonal matrix $\boldsymbol{K}$ such that $\boldsymbol{L}_s(\mathcal{G}(\boldsymbol{x}^0))= \boldsymbol{K}\boldsymbol{L}(\mathcal{G}(\boldsymbol{x}^0)) \boldsymbol{K}^{-1}$ is positive semidefinite. When the network is radial and the active power flow graph has positive weights, it is known that there exists a diagonal matrix $\boldsymbol{K}$ such that $\boldsymbol{L}_s(\mathcal{G}(\boldsymbol{x}^0))= \boldsymbol{K}\boldsymbol{L}(\mathcal{G}(\boldsymbol{x}^0)) \boldsymbol{K}^{-1}$ is symmetric. This follows from the standard results on diagonal symmetrizability of Metzler matrices whose associated graphs are radial \cite{norris1998markov}. Therefore, if the microgrid's active power flow graph is radial with positive edge weights, $\boldsymbol{L}_s(\mathcal{G}(\boldsymbol{x}^0))$ is guaranteed to be positive semidefinite with a non-repeated eigenvalue at zero. As mentioned before, for the lossy case, all  edge weights will be positive if the inequality $0 < ( \phi_{ik} \pm |\theta_i^0-\theta _k^0|)~ \mbox{mod}~ 2\pi < \pi$ is satisfied for all edges $(i,k)$ in the active power flow graph. For a radial network, the power flow equations can be shown to always have a solution where this condition is met.\footnote{Deletion of edge for which edge weight is zero from active power flow graph does not change the Laplacian and thus we assume there are no such lines.} This can be verified through a recursive argument, where the the angle for each leaf node (bus) is calculated first, and then the calculation is repeated recursively for remaining buses for which only one angle is unknown in the power flow equations. This analysis yields the following corollary:


\medskip
\noindent \textbf{Corollary 3:}  \textit{Consider The small-signal model (\ref{eq13}) for the lossy microgrid. The small-signal model is asymptotically stable for any filtering time constant, when the mircogrid is radial, and the active power flow graph does not have any critical lines. For a radial network, an equilibrium always exists such that the active power flow graph does not have any critical lines.}

\medskip
\noindent

\noindent \textbf{Discussion}: Illustrated results in Sections \ref{sec:Small-Signal Stability Analysis of Lossless Microgrid} and \ref{sec:Small disturbance stability analysis of lossy microgrid} are based on the known and fixed loads/generations. Microgrid planning under uncertainty in loads/generations is discussed widely from different aspects in current literature \cite{xiang2016robust,khodaei2015microgrid}. Uncertainty in loads/generation alter the operating solution, and hence the linearized model. The combined effect of this variability and lag/ filtering depends on the network structure. Specifically, in the lossless case, the impact of time-lag on stability is orthogonal to the impact of uncertainty, as the stability property is maintained regardless of the lag. Thus, if the no-lag model is stable over the range of operating uncertainties, the lagged model will be stable as well. In the lossy case, however, lags may alter the stability property, and hence there is a complex relationship between uncertainty and lag tolerance. We note that characterizing load/generation profiles which yield stability is an important (and difficult) question in its own right, even without considering lag/filtering, and it is out of the scope of this paper.


%


\section{Simulation Results}
\label{sec:simulation} 

In this section we aim to show simulation results which support our presented theoretical results. First we demonstrate simulation results showing the effect of lag/filtering on the stability and transient response of lossless microgrid. Next we present further results for lossy microgrids with both mesh and radial topologies. The simulations of transients are undertaken using the nonlinear DAE model developed in Section \ref{subsec:DAE Model for the Microgrid Network} if not mentioned otherwise.

\vspace{-0.1cm}
\subsection{Lossless Case}
\label{sec:Lossless Case}
\begin{figure}[h]
		\centering
		\vspace{-0.3cm}
		\includegraphics[width=0.8\linewidth]{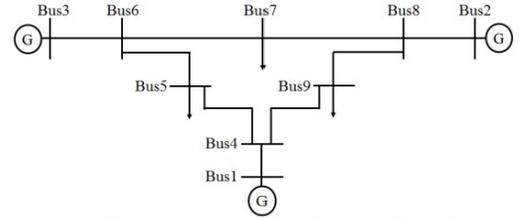}
		\vspace{-0.45cm}
		\caption{Diagram of the IEEE 9-Bus Test System.}
		\label{fig:2}
		\vspace{-0.3cm}
	\end{figure}

Here we demonstrate the effects of filtering on the transient response of a droop-controlled lossless microgrid. Specifically, we evaluate the effects of increasing lag on damping and settling time for the bus angles of microgrid. We first present the simulation results using modified IEEE 9-bus test system with the same model parameters as those used in \cite{song2015small}. In the modified test system (see Fig. \ref{fig:2}), buses 1, 2 and 3 are interfaced with inverter-based DERs (belong to ${\mathcal{V}}_A$), buses 5, 7 and 9 are non-inverter buses supplying for frequency dependent loads (belong to ${\mathcal{V}}_B$), and buses 4, 6 and 8 are supplying constant (i.e. zero) power loads (belong to ${\mathcal{V}}_C$). The bus and line parameters are detailed in Table \ref{table1} and Table \ref{table2}, respectively in the appendix section. 

The operation of the network is analyzed at two different equilibrium points: A and B. The bus angles obtained from power flow solution corresponding to both equilibrium points are shown in Table \ref{table2}. They admit the following characteristics:

{\em 1) Point A} is the normal operating point for the system. None of the lines satisfy the definition of a critical line. Therefore, for equilibrium point A, $\boldsymbol{L}(\mathcal{G}(\boldsymbol{x}^0))$ is positive semidefinite.

{\em 2) Point B} represents an unstable operating point for the test system. Corresponding to this point, lines $(5,6)$ and $(8,9)$ satisfy the definition for critical lines. It is found that $\boldsymbol{L}(\mathcal{G}(\boldsymbol{x}^0))$ is not positive semidefinite.
    \begin{figure}[t]
		\centering
		\includegraphics[width=3.1in]{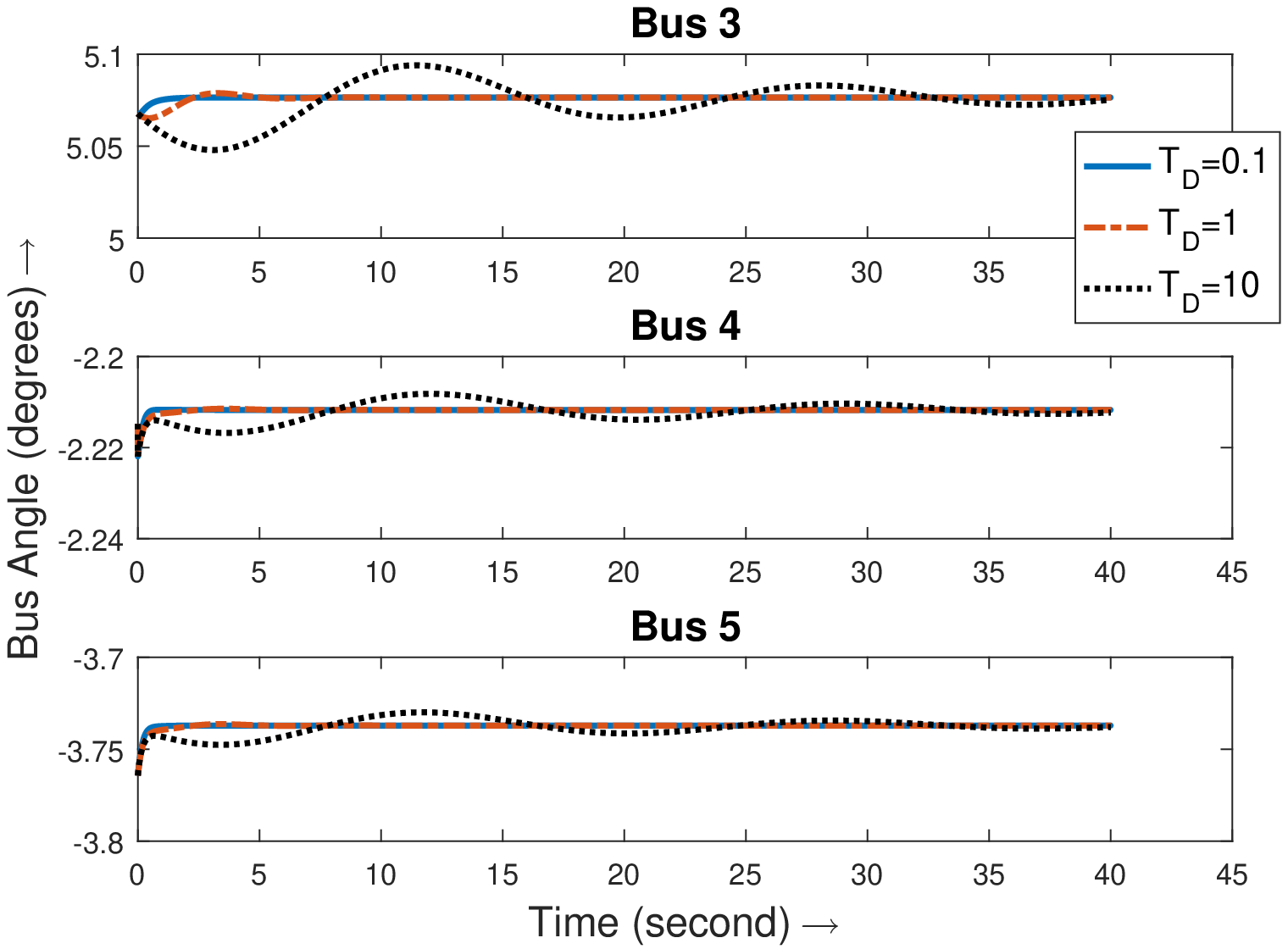}
			\vspace{-0.2cm}
        \includegraphics[width=3.1in]{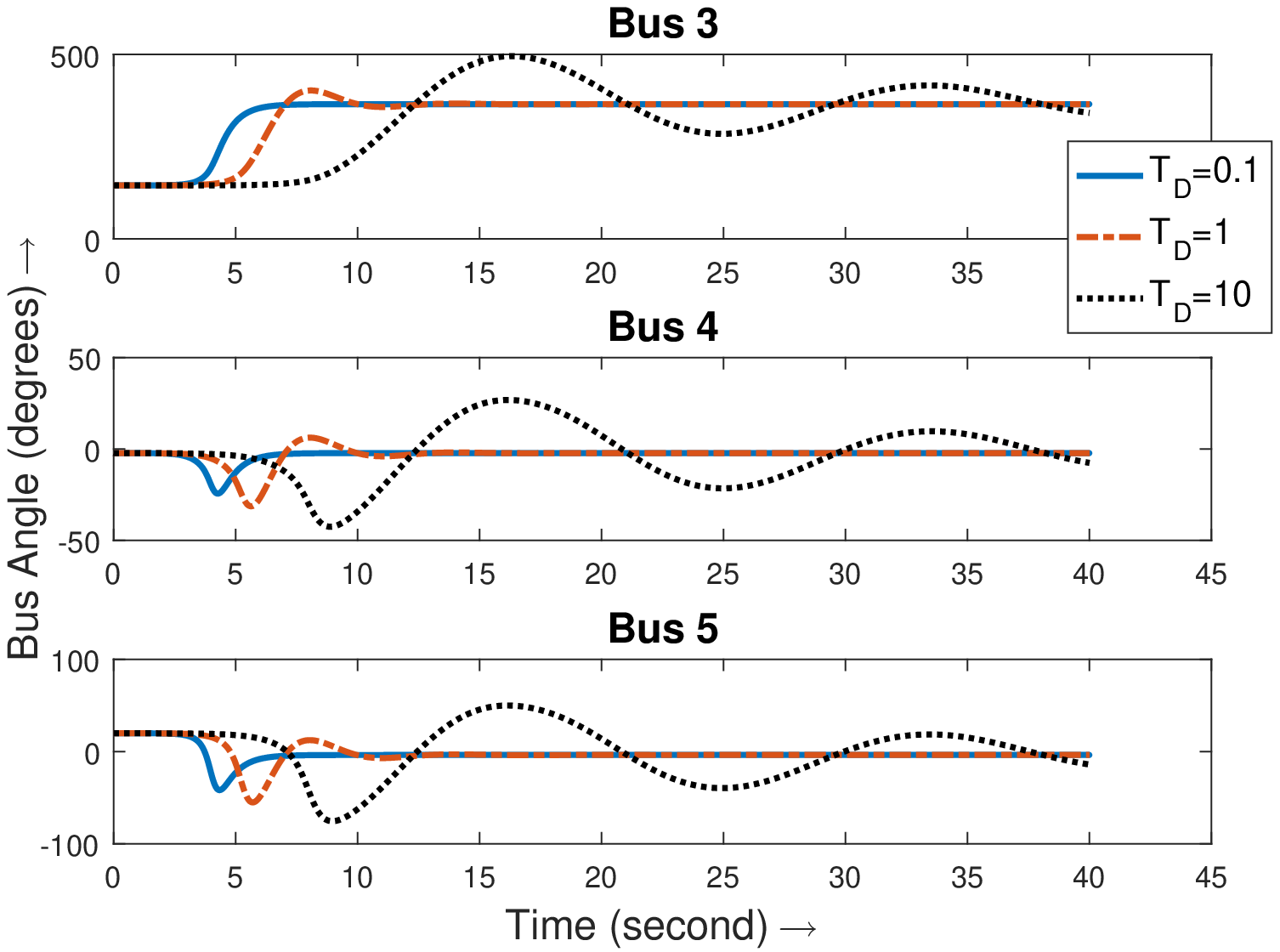}
        		\vspace{-0.2cm}
		\caption{Small-Disturbance Angle Response of Modified IEEE-9 Bus Lossless System for $T_D=$ 0.1, 1, 10 sec: (a) Point A (Small-Signal Stable); (b) Point B (Small-Signal Unstable)}
        \vspace{-0.3cm}
		\label{fig:3}
	\end{figure}
The angle response obtained from simulating transient behavior of test system due to small disturbances around both equilibrium points, A and B, are shown in Figs. \ref{fig:3}(a) and 3(b), respectively. The angle response are shown using three representative buses one from each category: inverter interfaced (bus 3), supplying frequency-dependent loads (bus 5), and supplying constant power loads (bus 4). The bus angles are plotted with respect to the angle measurement of bus 1. Note that the same time constant, $T_D$, is used for all the droop-controlled inverters. Three cases with different time constant ($T_D$ = 0.1, 1, and 10 sec) are simulated for both equilibrium points.

As it can be seen from Fig. \ref{fig:3}(a) the system is small-signal stable at point A regardless of the time constant of low pass filter of the droop controller. Similar conclusion is drawn from Theorem 2 as the Laplacian $\boldsymbol{L}(\mathcal{G}(\boldsymbol{x}^0))$ at point A is positive semidefinite with a nonrepeated eigenvalue at zero. Fig. \ref{fig:3}(a) also depicts the effect of lag on the settling time for microgrid. It is observed that the higher the time constant $T_D$, the longer the system takes to settle to the equilibrium point. Furthermore, buses having inverters, for example bus 3, shows a comparatively low damping and higher settling time compared to the load buses.

\begin{figure}[t]
		\centering
		\includegraphics[width=3.1in]{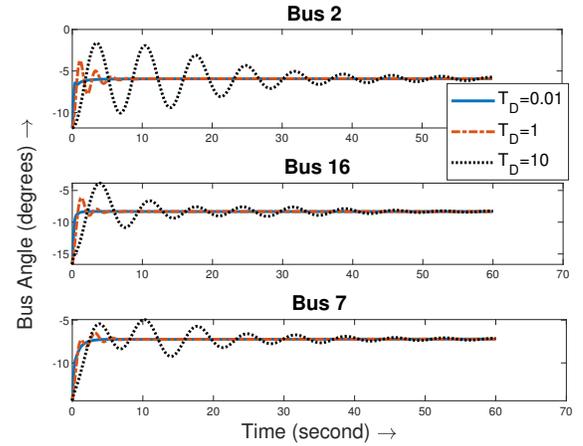}
        		\vspace{-0.2cm}
		\caption{Small-Disturbance Angle Response of Modified IEEE-57 bus system for $T_D=$ 0.1, 1, 10 sec}
		\label{fig:57bus}
	\end{figure}

It can be observed from Fig. \ref{fig:3}(b) that the other equilibrium, point B, is not small-signal stable. Since the corresponding $\boldsymbol{L}(\mathcal{G}(\boldsymbol{x}^0))$ is not positive semidefinite, the instability of point B is immediate from Theorem 2. The unstable behavior of point B can be further verified from the eigen analysis of the corresponding Jacobian matrix. When evaluated at point B, $\boldsymbol{J}(\boldsymbol{x}^0)$ admits one eigenvalue in the right half plane (i.e. $\lambda_{unstable}= 2.42$ for $T_D= 0.1$). Therefore, the microgrid admits unstable behavior for this point B, as shown in Fig. \ref{fig:3} (b).
	\begin{figure}[t]
	\vspace{-0.29cm}
		\centering
		\includegraphics[width=0.3\textwidth]{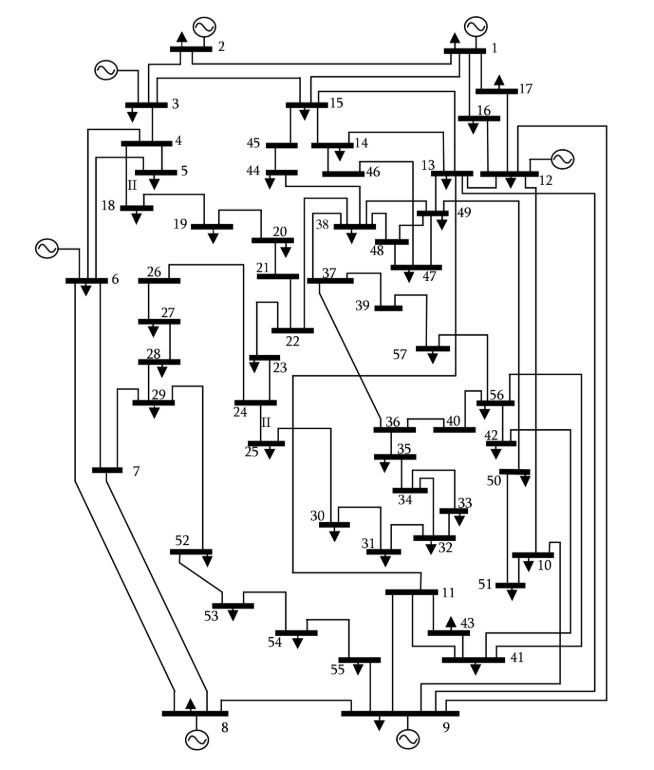}
		\vspace{-0.28cm}
		\caption{Diagram of the IEEE 57-Bus Test System.}
		\label{fig:57_diagram}
		\vspace{-0.28cm}
	\end{figure}
We also undertake simulations on a modified IEEE-57 bus system (see Fig. \ref{fig:57_diagram} for the network diagram) to show the effect of lag on the transient response (i.e. damping, settling time) of a larger network. We use the same parameter values given in \cite{christie1993ieee} but substitute the generator with inverter-based DERs, transformers with their equivalent reactances and ignore the line resistances according to our lossless assumption. We use $D_{R_i}=5$ and $D_{L_i}=2$ for inverters and frequency dependent loads, respectively. We obtain a stable equilibrium point using Newton-Raphson power flow algorithm (equilibrium is not shown due the space constraint) and simulate the angle response for the small disturbances around that equilibrium. Particularly for this simulation we utilize the small-signal model developed in Section \ref{subsec:Small-Signal Model} with $\epsilon_1=0.0001$, $\epsilon_2=0.01$. We consider each filter has the same time constant $T_D$. The angle response for buses $2$, $16$ and $7$, that belong to $\mathcal{V}_A$, $\mathcal{V}_B$ and $\mathcal{V}_C$, respectively are shown in in Fig. \ref{fig:57bus}. The angle response are shown for three different values of time constant, $T_D$, where angles  are measured with respect to angle at bus 1. Similar to the results of modified IEEE 9-bus system, we see that the system remains stable irrespective of time constant while damping decreases and settling time increases with the increase in the magnitude of the time constant. Also, as before, the inverter based buses, for example bus $2$, show low damping and higher settling time compared to the load buses.
 \vspace{-0.3cm}
\medskip
\subsection{Lossy Case}
In the subsequent subsections we show simulation results for lossy mesh and lossy radial microgrids. 
\subsubsection{Lossy Mesh Network}

\begin{figure}[t]
		\centering
		\includegraphics[width=0.35\textwidth]{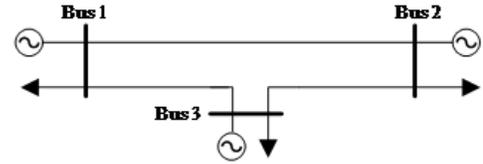}
		\vspace{-0.3cm}
		\caption{Diagram of the Considered 3 Bus Lossy Network.}
		\label{fig:lossy}
		\vspace{-0.5cm}
	\end{figure}
We present a simulation case study to verify the effects of filtering/lag for a lossy microgrid with mesh topology. A simple three-bus lossy microgrid shown in Fig. \ref{fig:lossy} is used for simulations. The bus and line parameters are detailed in Table \ref{table1_lossy} and \ref{table2_lossy} in the appendix section. Table \ref{table2_lossy} also includes an equilibrium point for the lossy three-bus system. We consider each filter has same time constant $T_D$. We show using simulations that the given system has a stable equilibrium for smaller time constant. However, for very large time constant, the system becomes unstable.

\begin{figure}[t]
		\centering
		\includegraphics[width=0.85\linewidth]{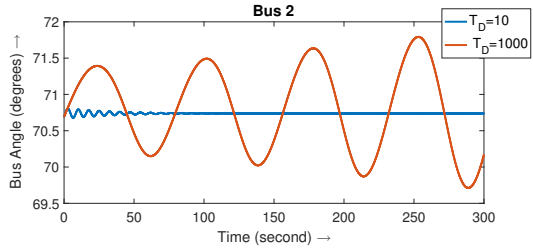}
        		\vspace{-0.3cm}
		\caption{Small-Disturbance Angle Response of Bus 2 of Considered Mesh Network for $T_D=$ 10 and 1000 sec.}
		\label{fig:lossy_res}
        \vspace{-0.5cm}
		\label{fig:5}
	\end{figure}

For the three-bus system, the Laplacian is given by: $\boldsymbol{L}(\mathcal{G}(\boldsymbol{x}^0))=$ $[0.183,-0.080,-0.103;-0.559,0.666,-0.106;-0.600,-0.033,0.634]$ which has all of its eigenvalue in the right half plane. Note that $\boldsymbol{W}(\boldsymbol{\theta}^0)$ is positive definite here and thus the system has no critical lines. Fig. \ref{fig:lossy_res} shows angle response of bus 2 for two different lags. It can be observed that for smaller lag i.e. $T_D=10s$, the angle reaches its equilibrium value. Whereas, for the larger lag, i.e. $T_D=1000s$, the system is unstable as angle diverges from the equilibrium point. This unstable behaviour can further be verified from the eigen-analysis of the corresponding Jacobian matrix. The $\boldsymbol{J}(\boldsymbol{x}^0)$ admits unstable eigenvalue $\lambda_{unstable}= 0.0002 \pm j0.0861$ for $T_D= 1000$s. In fact, according to Lemma 3, the system becomes unstable for any time constant that is greater than $522.7692$s. For large and highly resistive networks, this value can be significantly smaller and thus caution is required to avoid instability in such cases. This result emphasizes the critical requirements that need to be considered in filter designing problem. Specifically, the range of filtering time constants should be chose carefully to ensure a stable system operation for a lossy microgrid. In addition, the effects of filtering time constant on damping and settling times of the microgrid should be taken into account. The filter design problem, however, is outside the scope of this paper.

	\vspace{-0.2cm}
\subsubsection{Lossy Radial Network}

\begin{figure}[t]\vspace{-0.4cm}
		\centering
		\includegraphics[width=3.1in]{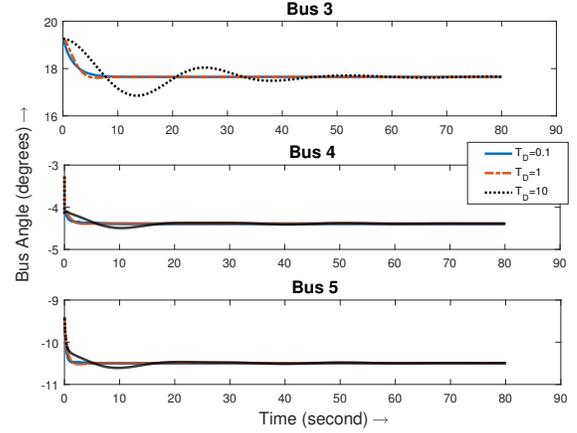}
        		\vspace{-0.3cm}
		\caption{Small-Disturbance Angle Response of Modified IEEE-9 Bus Lossy System with Radial Topology for $T_D=$0.01, 1 and 10 sec.}
        \vspace{-0.6cm}
		\label{fig:treelossy}
	\end{figure}

Here, we investigate the effects of filtering on small-signal stability of a lossy radial mircrogrid. We obtain a radial microgrid  by removing the line from bus 5 to bus 6 in Fig. \ref{fig:2}. Here we consider each line to be lossy with $R/X$ ratio as 0.5.  All the load and generation parameters for buses in the obtained radial system are same as the ones in Table \ref{table1} except the bus 1 whose active power generation changed to $1.0831$ p.u to compensate the ohm losses of the network. The line parameters and the equilibrium point are given in Table \ref{table 5} in the appendix section. We consider each filter has same time constant $T_D$ as before. It can be seen in Fig. \ref{fig:treelossy} that the equilibrium is small-signal stable regardless of the time constant of the low pass filter in the droop control. It is immediate from Corollary 3 as the network is radial and its corresponding active power flow graph has positive edge weights (equivalently, does not have any critical lines). Note that, there exists a diagonal matrix $\boldsymbol{K}=diag(0.5958, 0.5138, 0.5089, 0.6191, 0.6532, 0.5248, 0.5523$, $0.5469, 0.6222)$ for which $\boldsymbol{L}_s(\mathcal{G}(\boldsymbol{x}^0))= \boldsymbol{K}\boldsymbol{L}(\mathcal{G}(\boldsymbol{x}^0)) \boldsymbol{K}^{-1}$ is positive semidefinite with a non-repeated eigenvalue at zero. Therefore, stability of the equilibrium is also guaranteed by Theorem 4.

The simulation results demonstrated here verify the theoretical results obtained in Section \ref{sec:Small disturbance stability analysis}. From simulations, the following are the key observations for a droop controller with filtering : (1) For lossless connected microgrid, lag/filtering can not destabilize the system irrespective of its topology; (2) An increased lag (due to increase of filter time constant) results in less damping and increases the settling time for the transient response of the bus angles; (3) The effects of lag damping and settling time for bus angles are more pronounced for generator buses; (4) For a lossy mesh microgrid, a large lag can destabilize the otherwise stable equilibrium point of the system, (5) For a lossy radial microgrid, larger lag cannot destabilize the system but affects the damping and settling time.
\vspace{-0.1cm}	
     \section{Conclusion}
     \label{sec:Conclusion}
In this paper, we thoroughly investigated the effects of lag/ filtering in frequency-droop control on the angle stability of inverter-based lossless and lossy microgrids. We developed a nonlinear DAE model and a structure preserving linear differential-equation approximate model for a microgrid with low-pass-filtered droop control. Using the proposed models, we derived necessary and sufficient condition for lossless microgrid that guarantees a stable operation regardless of lag in droop control and also discussed graph-theoretic implications. Simulation case studies are presented using IEEE 9-bus and 57-bus test systems to demonstrate the transient response of the lossless microgrid with lagged droop-control. It is observed that on increasing lag the system takes longer time to settle to the equilibrium point. For the microgrid with lossy lines, sufficient conditions have been derived when there are no lag/filtering in the droop control and later by providing a condition we prove that the previously derived sufficient conditions do not apply in general when lag/filtering is present and instability may result on increasing the lag. A simulation case study is also included for a 3-bus lossy microgrid that demonstrates the destabilization of a stable equilibrium system on increasing the lag.  Further, for a tree/radial lossy microgrid, sufficient condition for stability is derived. It is demonstrated using modified IEEE 9-bus test system that for a radial microgrid, the droop-control with lag/filtering does not affect the otherwise stable equilibrium point, however, damping and settling times are affected. These crucial observations caution us against generalizing the stability conditions obtained using structure-preserving models for a practical lossy microgrid that usually admits a time-lagged droop response.
     
Future research directions include the analysis of the voltage dynamics and consequently the effects of $Q-droop$ on microgrid's small signal stability. Another important future direction is to study the impacts of uncertainty in loads/generation on the small-signal stability of the microgrid. Finally, it would be interesting to improve the stability of the microgrid by designing the droop controllers' gain. This could be formulated as an optimization problem where the objective could be to optimize a stability metric (e.g. an optimal damping ratio) given bounds on the lag introduced due to filtering. Finally, although our focus here is on a primary control, overlaid secondary and tertiary controllers for microgrids depend on longer-range communications, and hence are also critically impacted by time-lags; co-design of primary and secondary controls in a way that accounts for the lags is an important direction of future work.  
    
    
\appendix

\section*{APPENDIX}

\section{Test systems}

Here we detail the parameter values of the test systems that we have used in our simulation in the following tables. 

	\begin{table}[thpb]
		\caption{Bus Generation and Load Parameters of Modified IEEE-9 Bus Lossless Test System}
        \vspace{-0.3cm}
		\label{table1}
		\begin{tabular}{cccccl}
			\toprule
						Bus & $P^0_{G_i}$ (p.u.) & $P^0_{L_i}$ (p.u.) & $V_i$(p.u.) & $D_{R_i}$(s)& $D_{L_i}$(s)\\
            \midrule
			1 &0.67&0&1&5&0\\
			2 &1.63&0&1&5&0\\
            3 &0.85&0&1&5&0\\
            4 &0&0&1&0&0\\
            5 &0&0.90&1&0&2\\
            6 &0&0&1&0&0\\
            7 &0&1.00&1&0&2\\
            8 &0&0&1&0&0\\
            9 &0&1.25&1&0&2\\
			\bottomrule
		\end{tabular}
	\end{table}
	
    \begin{table}[thpb]
		\caption{Line Parameters and Angle Differences across Lines at
Equilibrium Point A and Point B for Modified IEEE-9 Bus Lossless Test System}
		\label{table2}
		\begin{tabular}{cccc}
			\toprule
						 Line&$X$ (p.u.) & $\theta_{i}-\theta_{k}$ at A  & $\theta_{i}-\theta_{k}$ at B\\
            \midrule
			(1, 4) &0.0576&2.21$^{\circ}$ &2.21$^{\circ}$\\
			(4, 5) &0.0920&1.53$^{\circ}$&-22.04$^{\circ}$\\
            (5, 6) &0.1700&-5.96$^{\circ}$&-122.17$^{\circ}$\\
            (3, 6) &0.0586&2.86$^{\circ}$&2.86$^{\circ}$\\
            (6, 7) &0.1008&1.38$^{\circ}$&-24.60$^{\circ}$\\
            (7, 8) &0.0720&-3.14$^{\circ}$&338.33$^{\circ}$\\
            (8, 2) &0.0625&-5.85$^{\circ}$&-5.85$^{\circ}$\\
            (8, 9) &0.1610&8.05$^{\circ}$&-145.71$^{\circ}$\\
            (9, 4) &0.0850&-1.85$^{\circ}$&-23.81$^{\circ}$\\
			\bottomrule
		\end{tabular}
	\end{table}
	
			\begin{table}[thpb]
		\caption{Bus Generation and Load Parameters of 3-Bus Lossy System}
		\label{table1_lossy}
		\begin{tabular}{cccccc}
		\toprule
						Bus & $P^0_{G_i}$ (p.u.) & $P^0_{L_i}$ (p.u.) & $V_i$(p.u.) & $D_{R_i}$(s)& $D_{L_i}$(s)\\
            \midrule
			1 &0.5295&2&1&0.1&0\\
			2 &3.0860&2&1&0.1&0\\
            3 &3.0650&2&1&0.1&0\\
			\bottomrule
		\end{tabular}
	\end{table}

        \begin{table}[thpb]
		\caption{Line Parameters and Angle Differences across Lines at Equilibrium Point of 3-Bus Lossy System}
		\label{table2_lossy}
		\begin{tabular}{ccc}
		\toprule
						 Line &$R+jX$ (p.u.) & $\theta_{i}-\theta_{k}$ \\
            \midrule
			(1, 2) & $0.254+j0.967$ &-70.7$^{\circ}$ \\
			(1, 3) & $0.267+j0.964$ &-68.6$^{\circ}$\\
            (2, 3) & $0.998+j0.070$ & 2.1$^{\circ}$\\
	\bottomrule
		\end{tabular}
	\end{table} 

	\begin{table}[thpb]
		\caption{Line Parameters and Angle Differences across Lines at
Equilibrium Point for Modified IEEE-9 Bus Lossy System with Radial Topology }
		\label{table 5}
		\begin{tabular}{ccc}
			\toprule
						 Line&$R+jX$ (p.u.) & $\theta_{i}-\theta_{k}$  \\
            \midrule
			(1, 4) &$0.0288+j0.0576$& 4.3886$^{\circ}$ \\
			(4, 5) &$0.0460+j0.0920$&6.1041$^{\circ}$\\
            (3, 6) &$0.0293+j0.0586$&3.5157$^{\circ}$\\
            (6, 7) &$0.0504+j0.1008$&5.8135$^{\circ}$\\
            (7, 8) &$0.0360+j0.0720$&-1.1227$^{\circ}$\\
            (8, 2) &$ 0.0313 +j0.0625$&-7.0950$^{\circ}$\\
            (8, 9) &$ 0.0805 + j0.1610$&14.3953$^{\circ}$\\
            (9, 4) &$ 0.0425+j0.0850$&-0.5655$^{\circ}$\\
			\bottomrule
		\end{tabular}
	\end{table}    

\bibliographystyle{ACM-Reference-Format}
\bibliography{sample-base}

\end{document}